\documentclass[aps,prl,twocolumn,superscriptaddress]{revtex4}
\usepackage{epsf,graphicx}
\usepackage{amssymb}
\usepackage{amsmath}
\usepackage{bm}

\usepackage{color}
\usepackage{ulem}
\begin{document}
\title{Development of long-range phase coherence on the Kondo lattice}
\author{Jian-Jun Dong}
\affiliation{Department of Physics and Chongqing Key Laboratory for Strongly Coupled Physics, Chongqing University, Chongqing 401331, China}
\affiliation{Beijing National Laboratory for Condensed Matter Physics and Institute of Physics, Chinese Academy of Sciences, Beijing 100190, China}
\affiliation{University of Chinese Academy of Sciences, Beijing 100049, China}
\author{Yi-feng Yang}
\email[]{yifeng@iphy.ac.cn}
\affiliation{Beijing National Laboratory for Condensed Matter Physics and Institute of Physics, Chinese Academy of Sciences, Beijing 100190, China}
\affiliation{University of Chinese Academy of Sciences, Beijing 100049, China}
\affiliation{Songshan Lake Materials Laboratory, Dongguan, Guangdong 523808, China}
\date{\today}

\begin{abstract}
Despite of many efforts, we still lack a clear picture on how heavy electrons emerge and develop on the Kondo lattice. Here we introduce a key concept named the hybridization bond phase and propose a scenario based on phase correlation to address this issue. The bond phase is a gauge-invariant quantity combining two onsite hybridization fields mediated by inter-site magnetic correlations. Its probabilistic distribution decays exponentially with site distance, from which a characteristic length scale can be extracted to describe the spatial correlation of Kondo hybridizations. Our calculations show that this correlation length grows logarithmically with lowering temperature at large Kondo coupling, and reveal a precursor pseudogap state with short-range phase correlation before long-range phase coherence is developed to form the Kondo insulating (or heavy electron) state at low temperatures. This provides a potential microscopic explanation of the two-stage hybridization proposed by recent pump-probe experiment and the logarithmic scaling in the phenomenological two-fluid model. Our work offers a novel theoretical framework to describe the phase-related physics in Kondo lattice systems.
\end{abstract}

\maketitle

Heavy fermion systems are featured with correlated phenomena such as unconventional superconductivity, quantum criticality, and non-Fermi liquid \cite{Sigrist1991RMP,Stewart2001RMP,Gegenwart2008NatPhys,Yang2012PNAS}. Underlying all  these exotic properties is the interplay of inter-site magnetic correlations and spin screening as described by the Kondo lattice model \cite{Hewson1997,Coleman2015,Yang2017PNAS}. It differs from the single-impurity Kondo physics in that the many-body spin-entangled state between local moments and conduction electrons can extend and propagate on the lattice as dispersive heavy electrons. But how this state emerges, develops and eventually extends in space remains less understood despite many theoretical efforts \cite{Lonzarich2017RPP}. In particular, recent angle-resolved photoemission spectroscopy (ARPES) experiment reported the onset of hybridization well above the coherence temperature in transport measurements \cite{Chen2017PRB,Yang2016RPP}. Later pump probe experiment revealed a two-stage process for the heavy electron development on the lattice \cite{Hu2019PRB,Liu2020PRL}. This two-stage scenario seems quite universal \cite{Pei2021PRB}, but still awaits a microscopic explanation. Its theoretical formulation will necessarily deepen our understanding of the Kondo lattice physics.

Phase fluctuations play an indispensable role in modern strongly correlated physics and have been extensively studied in past decades. Notable examples include deconfined phases in lattice gauge theories \cite{Kogut1979RMP,Gazit2017NatPhys,Xu2019PRX}, phase-fluctuation scenario in high-$T_c$ cuprates  \cite{Dagotto2005PRL,Dagotto2008PRL,Dubi2007Nature,TaoLi2010PRB}, and flux phases in quantum spin liquids \cite{Wen2004,Zhou2017RMP,Wang2021PRL}. In Kondo lattice systems, phase fluctuations appear when some slave particles or auxiliary fields are introduced to describe local spins and their entanglement with conduction electrons, but were often ignored in prevailing mean-field calculations, causing artificial finite-temperature phase transitions \cite{Auerbach1986PRL,Newns1987AdvPhys,Zhang2000PRB,Coqblin2003PRB,Pepin2011PRL,Zhang2018PRB}. Phase fluctuations can convert these artificial transitions into a crossover \cite{Senthil2003PRL,Senthil2004PRB}, and may be a key controlling the development of heavy electron state beyond usual mean-field or local approximations  \cite{Coleman2005PRB,Pepin2007PRL,Pepin2008PRB,Ohara2007PRB,Ohara2013JPSJ,Ohara2014JPSJ,Jiang2017PRB,Jiang2020PRB,Hu2020PRR,Han2021PRB}, but a proper description is not yet available.

In this work, we address this issue by proposing a novel theoretical framework to describe heavy electron emergence based on phase coherence. We employ a recently-proposed static auxiliary field approach for Kondo systems \cite{Dong2021PRB} and introduce a gauge-invariant bond phase to characterize spatial correlation of the hybridization fields. This bond phase combines the onsite hybridization fields and inter-site magnetic correlation fields between two spatially separated lattice sites, so its probabilistic distribution tracks directly the spatial development of heavy electron state. Our calculations show a logarithmic temperature dependence of its characteristic length scale and reveal a precursor short-range phase-correlated pseudogap state before the long-range phase coherence is developed to form the Kondo insulating or heavy electron state at lower temperatures. Our theory provides a general scheme to describe the phase coherence and other phase-related physics in Kondo lattice systems.

For simplicity, we consider the two-dimensional Kondo-Heisenberg model on a square lattice,
\begin{equation}
H=-t\sum_{\left\langle ij\right\rangle \sigma}\left(  c_{i\sigma}^{\dag}c_{j\sigma}+\text{H.c.}\right)  +J_{\text{K}}\sum_{i}\mathbf{s}_{i} \cdot\mathbf{S}_{i}+J_{\text{H}}\sum_{\left\langle ij\right\rangle } \mathbf{S}_{i}\cdot\mathbf{S}_{j},
\end{equation}
where $t$ is the hopping integral of conduction electrons between nearest-neighbor sites, $\mathbf{s}_{i}=\sum_{a\beta}c_{i\alpha}^{\dag}\frac{\bm{\sigma}_{\alpha\beta}}{2}c_{i\beta}$ is the conduction electron spin localized at $\mathbf{r}_{i}$, and $\mathbf{S}_{i}$ denotes the local spins. $J_{\text{K}}$ and $J_{\text{H}}$ describe the Kondo and Heisenberg exchange interactions. Under the Abrikosov pseudofermion representation $\mathbf{S}_{i}=\sum_{\eta\gamma}f_{i\eta}^{\dag}\frac{\bm{\sigma}_{\eta\gamma}}{2}f_{i\gamma}$, both terms can be decoupled using the Hubbard-Stratonovich transformation: $2\mathbf{s}_{i}\cdot\mathbf{S}_{i}  \rightarrow \sum_{\sigma}(V_{i}c_{i\sigma}^{\dag}f_{i\sigma}+\text{H.c.})+\left\vert V_{i}\right\vert^{2}$ and $2\mathbf{S}_{i}\cdot\mathbf{S}_{j}  \rightarrow\sum_{\sigma}(\chi_{ij}f_{i\sigma}^{\dag}f_{j\sigma}+\text{H.c.})+\left\vert\chi_{ij}\right\vert ^{2}$, where $V_i$ and $\chi_{ij}$ are two fluctuating auxiliary fields describing the hybridization and inter-site magnetic correlation, respectively. This gives the action \cite{SM}:
\begin{align}
S  &  =\sum_{i,l}\frac{\beta J_{\text{K}}\left\vert V_{i,l}\right\vert ^{2}}{2}+\sum_{\left\langle ij\right\rangle ,l}\frac{\beta J_{\text{H}}\left\vert \chi_{ij,l}\right\vert ^{2}}{2}-\sum_{i}\beta\lambda_{i}\nonumber\\
&  \quad+\sum_{nm\sigma}\Psi_{n\sigma}^{\dag}\left(  O_{nm}-\operatorname*{i} \omega_{n}\delta_{nm}\right)  \Psi_{m\sigma},
\label{eq2}
\end{align}
where $\Psi_{n\sigma}=\left[c_{1\sigma n},\cdots,c_{N_{0}\sigma n},f_{1\sigma n},\cdots,f_{N_{0}\sigma n}\right]  ^{T}$, $N_{0}$ is the number of lattice sites, and the subscripts $l$ ($n/m$) denote bosonic (fermionic) Matsubara frequency. $\lambda_{i}$ is the Lagrange multiplier for the constraint $\sum_{\sigma}f_{i\sigma}^{\dag}f_{i\sigma}=1$ and takes a real value after Wick rotation \cite{Zhou2017RMP}.

The above action is generally impossible to solve. To proceed, we adopt a static approximation, $V_{i,n-m}=V_{i}\delta_{nm}$, $\chi_{ij,n-m}=\chi_{ij}\delta_{nm}$ such that $O_{nm}=O\delta_{nm}$. This ignores temporal fluctuations of the auxiliary fields but takes full account of their spatial fluctuations and statistical distribution \cite{Mukherjee2014PRB,Pradhan2015PRB,Karmakar2016PRA,Patel2017PRL}. The fermions can be integrated out, giving an effective action only of the auxiliary fields:
\begin{align}
S_{\text{eff}}  &  =\sum_{i}\frac{\beta J_{\text{K}}\left\vert V_{i} \right\vert ^{2}}{2}+\sum_{\left\langle ij\right\rangle }\frac{\beta J_{\text{H}}\left\vert \chi_{ij}\right\vert ^{2}}{2}-\sum_{i}\beta\lambda_{i}\nonumber\\
&  \quad-2\sum_{n}\ln\det\left(  O-\operatorname*{i}\omega_{n}\right). \label{effective-action}
\end{align}
The matrix $O$ has a block form $O=\left[ \begin{array} [c]{cc} T^{c} & M \\ M^{\dag} & T^{f} \end{array} \right]$, where $T^{c}$ ($T^{f}$) is the $N_{0}\times N_{0}$ hopping matrix of conduction electrons (pseudofermions), and $M_{ij}=\delta_{ij}J_{\text{K}}V_i/2$ is a diagonal matrix for their onsite hybridization. The summation over Matsubara frequency can be evaluated using $\sum_{n}\ln \det\left(  O-\operatorname*{i}\omega_{n}\right)  =\sum_{l}\ln\left(1+\operatorname{e}^{-\beta\xi_{l}}\right)$, where $\xi_{l}$ is the eigenvalues of $O$ and always real because $O$ is Hermitian. The probabilistic distribution of the auxiliary fields is then simply, $p(V_{i},\chi_{ij})=Z^{-1}\operatorname{e}^{-S_{\text{eff}}}$, where $Z$ is the partition function serving as the normalization factor. It can be simulated using the Monte Carlo and Metropolis algorithm on $3N_0$ complex random variables without sign problem \cite{Ishizuka2012PRL,Yang2019PRB,Maska2020PRB}. This is different from the (non-uniform) mean-field method where the auxiliary variables take fixed values determined by the saddle-point approximation. For simplicity, we set the half conduction bandwidth to unity ($t=1/4$), fix $J_{\text{H}}=0.2$, and consider only the particle-hole symmetric model on a $N_{0}=8\times8$ lattice where the Lagrangian multipliers are approximated by their saddle-point value $\lambda_{i}=0$ \cite{Saremi2011PRB}. Other choices of parameters or a larger lattice have been examined and the conclusions are qualitatively unchanged.

\begin{figure}[ptb]
\begin{center}
\includegraphics[width=8.6cm]{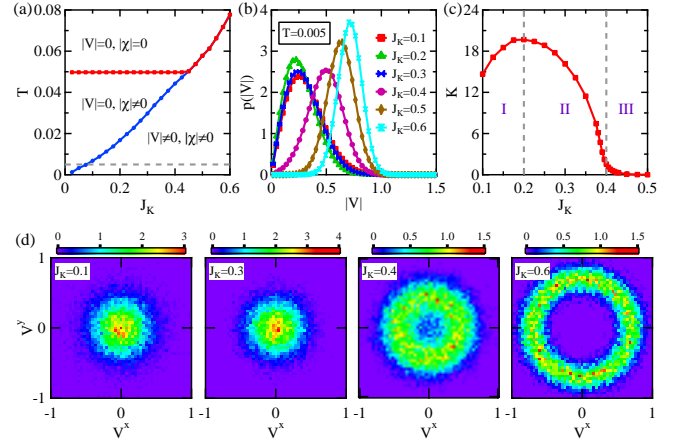}
\end{center}
\caption{(a) The mean-field phase diagram predicted under uniform approximation. The dashed line marks the temperature $T=0.005$. (b) The amplitude probabilistic distribution $p (  |V| )$ for different $J_\text{K}$ at $T=0.005$. (c) The slope $K=dp(  \left\vert V\right\vert )  /d\left\vert V\right\vert$ at $\left\vert V\right\vert=0$ showing three different regions. (d) Density plots of $p (  V )$ on complex plane $V=(V^x,V^y )$ for four chosen values of $J_{\text{K}}$ at $T= 0.005$.
}
\label{fig1}%
\end{figure}

For comparison, we first show the usual mean-field phase diagram in Fig.~\ref{fig1}(a), where the auxiliary fields are assumed to be uniform and real: $V_{i}=\overline{V}_{i}=V$, $\chi_{ij}=\overline{\chi}_{ij}=\chi$. The solution can be obtained by minimizing the free energy $F=S_{\text{eff}}/\beta$ \cite{SM}. With lowering temperature, we see a second-order phase transition from $\chi =0$ to $ \chi \neq0$ at $T=J_{\text{H}}/4$, below which there is a weakly coupled state of conduction electrons and spin liquid \cite{Senthil2003PRL}. Increasing $J_{\text{K}}$ drives the system into a Kondo insulating state with nonzero $V$.

We may examine this mean-field picture by considering the probabilistic distribution of the complex hybridization fields, $p(V)\equiv p(V_i)$, which is the same on all sites due to translational symmetry and can be evaluated using the Metropolis algorithm for importance sampling of the effective action Eq.~(\ref{effective-action}) \cite{SM}. Figure~\ref{fig1}(b) plots the marginal distribution of its amplitude, $p(  \vert V\vert )  $, at a low temperature after integrating out all other variables. The maximum of $p(\vert V\vert )  $ is seen to vary nonmonotonically with increasing $J_{\text{K}}$. As shown in Fig.~\ref{fig1}(c), we may identify three regions according to the slope $K=dp(  \left\vert V\right\vert )  /d\left\vert V\right\vert$ at $\left\vert V\right\vert=0$. Figure~\ref{fig1}(d) plots the distribution $p (  V)$ on the complex plane $V=\left(V^x,V^y \right)$. As expected, the data cluster around $(  0,0 )$ at small $J_{\text{K}}$ and turn into a ring at large $J_{\text{K}}$. The distribution is therefore dominated by the bare fluctuation term $\beta J_K|V|^2/2$ in region I and the coupling with conduction electrons in region III, while region II marks a crossover in between.

\begin{figure}[ptb]
\begin{center}
\includegraphics[width=8.6cm]{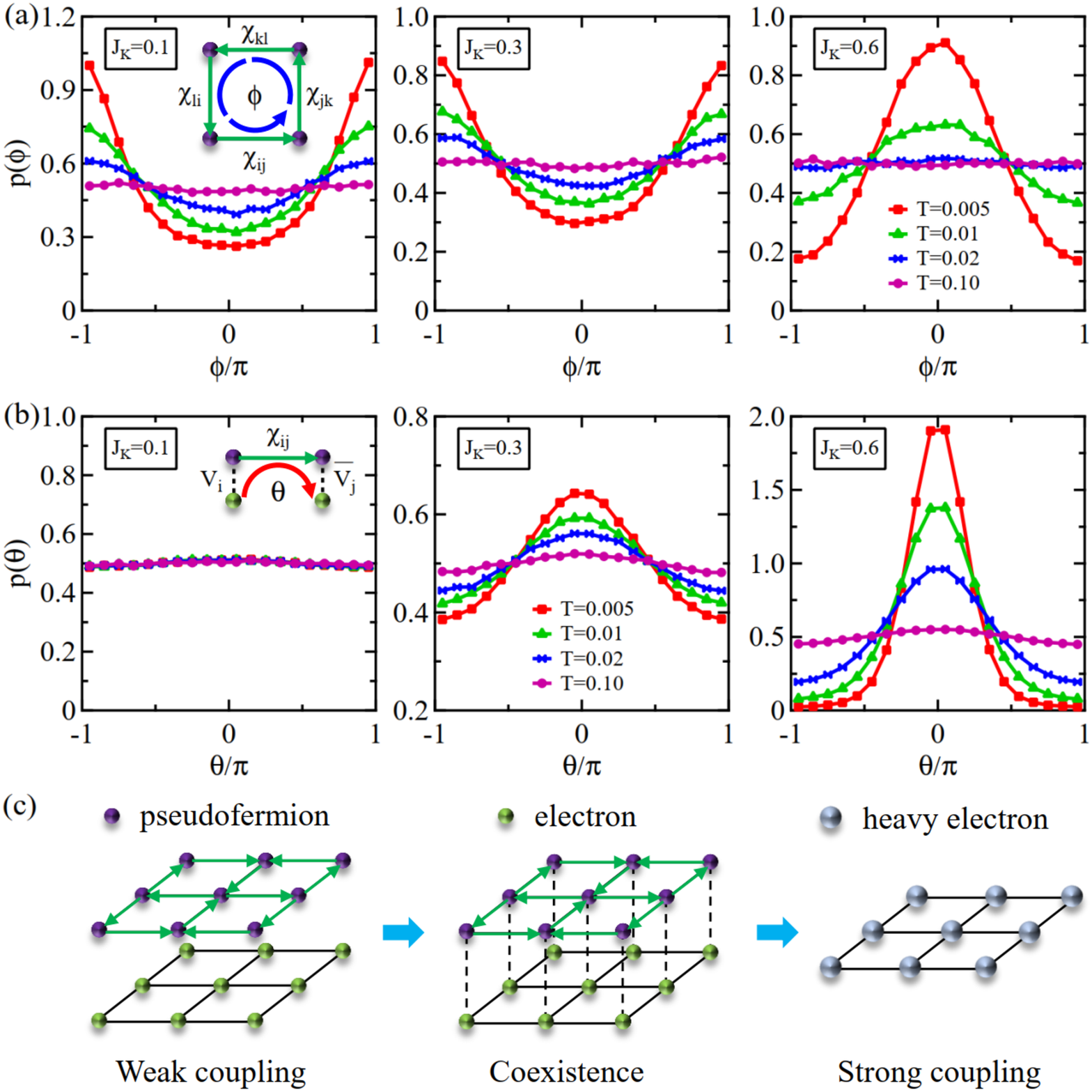}
\end{center}
\caption{Comparison of the probabilistic distribution of (a) the flux $p(\phi)$ and (b) the bond phase $p(\theta)$ for three typical values of $J_{\text{K}}$. The insets illustrate the definition of flux and bond phases. (c) Schematic plots of the coupling between pseudofermions and conduction electrons in three different low-temperature regions.}
\label{fig2}
\end{figure}

The difference from the mean-field solution can be revealed by studying phase fluctuations of the auxiliary fields. Since the effective action (\ref{effective-action}) is invariant under the gauge transformation $V_{i}\rightarrow V_{i}\operatorname{e}^{\operatorname*{i}\beta_{i}}$, $\chi_{ij}\rightarrow\operatorname{e}^{-\operatorname*{i}\left(  \beta _{i}-\beta_{j}\right)  }$, we may define two gauge-invariant phases from
\begin{align}
F_{i}  & \equiv\chi_{ij}\chi_{jk}\chi_{kl}\chi_{li}=\left\vert F_{i}%
\right\vert \operatorname{e}^{\operatorname*{i}\phi_{i}},\nonumber\\
B_{ij}  & \equiv V_{i}\chi_{ij}\overline{V}_{j}=\left\vert B_{ij}\right\vert
\operatorname{e}^{\operatorname*{i}\theta_{ij}},
\end{align}
where $\phi_{i}$ denotes the flux in a plaquette $ ijkl \in \Box$ and $\theta_{ij}$ reflects the phase of the hybridization bond $B_{ij}$ between nearest-neighbor sites $ij$ as illustrated in the insets of Fig.~\ref{fig2}. The bond phase $\theta_{ij}$ reflects the  correlation of two onsite hybridization fields $V_i$ and $V_j$ mediated by their inter-site magnetic correlation $\chi_{ij}$. It by definition contains spatial correlation information and is a key quantity introduced in this work to distinguish the Kondo lattice physics from the single-impurity Kondo physics \cite{Nakatsuji2002PRL,Yang2008Nature,Wang2021PRB,Wang2022SCPMA}.

\begin{figure}[ptb]
\begin{center}
\includegraphics[width=8.6cm]{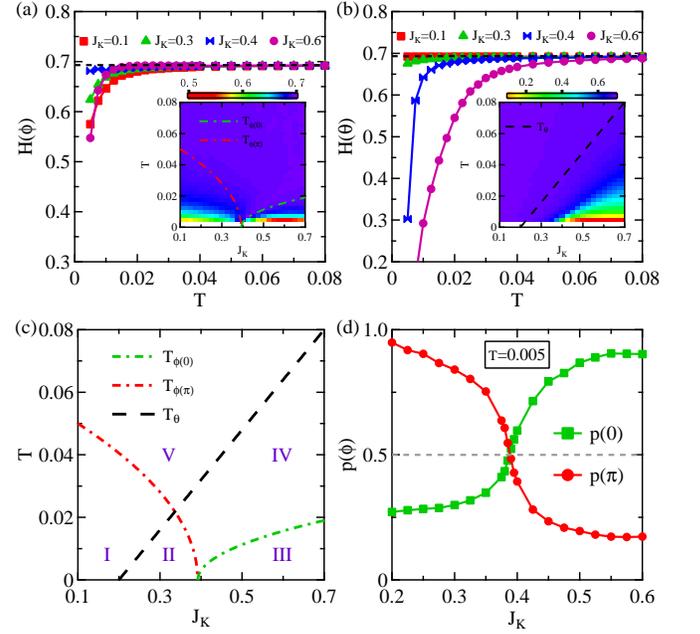}
\end{center}
\caption{The Shannon entropy of (a) the flux $\phi/\pi$ and (b) the bond phase $\theta/\pi$ as functions of temperature for different values of $J_{\text{K}}$. The black dashed line marks the uniform limit $\ln 2$. The insets are the intensity plot of the Shannon entropy on the $T$-$J_{\text{K}}$ plane, showing different behaviors of flux and bond phases at low temperatures. The lines are a guide to the eye separating the diagram into different tentative regions. (c) A schematic $T$-$J_{\text{K}}$ phase diagram constructed based on the intensity plots, where $T_\theta$ marks a crossover between uniform and non-uniform distribution of the bond phase, and $T_{\phi(0)}$ ($T_{\phi(\pi)}$) marks a crossover to the non-uniform flux distribution peaked at $\phi=0$ ($\pi$). (d) Comparison of the flux probabilistic distribution $p(\phi)$ at $\phi=0$ and $\pi$  as a function of the Kondo coupling at a very low temperature $T=0.005$.}
\label{fig3}
\end{figure}

Figures~\ref{fig2}(a) and \ref{fig2}(b) plot their probabilistic distributions for three typical values of $J_{\text{K}}$. Again, due to translational symmetry, we drop the site subscript and use  $p(\phi)\equiv p(\phi_{i})$ and $p(\theta)\equiv p(\theta_{ij})$. For weak coupling $J_{\text{K}}=0.1$, the local spins and conduction electrons are almost decoupled. The bond phase $p(\theta)$ distributes uniformly at all temperatures, while the flux distribution $p(\phi)$ becomes peaked at $\phi=\pi$ below certain temperature. The latter indicates a $\pi$-flux state  \cite{Affleck1988PRB,Hsu1990PRB,Wen2002PRB,Coleman2021PRR}, which is a special feature of the pseudofermion representation of the Heisenberg model on the two-dimensional lattice and might be realized if the spin interaction is highly frustrated or on an optical lattice. Our results are in accordance with Lieb's theorem \cite{Lieb1994PRL}, which states that the saddle point for a half-filled band of fermions hopping on a planar lattice is $\pi$ per plaquette. This implies that our approach can capture the correct saddle point beyond the ad hoc uniform mean-field approximation. The $\pi$-flux state may not be stable in general situations, giving rise to confined states like antiferromagnetic order or magnon excitations. For simplicity, we ignore all these complications and take it as our starting point in order to focus on the Kondo aspect of the model. For an intermediate $J_{\text{K}}=0.3$, we still have $\pi$-flux, but the bond phase distribution becomes non-uniform at low temperatures with a peak around $\theta=0$, indicating the onset of phase correlation between nearest-neighbor hybridization fields. It therefore marks a coexisting state of flux and hybridization correlation. For strong coupling $J_{\text{K}}=0.6$, the location of maximal $p(\phi)$ changes from $\phi=\pi$ to $\phi=0$, indicating that the $\pi$-flux is completely suppressed and the system enters the Kondo insulating (or heavy electron) state. Figure~\ref{fig2}(c) gives an illustration of all three low-temperature states. For completeness, the probabilistic distributions of $|F_i|$ and $|B_{ij}|$ are also given in the Supplemental Material \cite{SM}.

The variation of the probabilistic distributions may be reflected also in their Shannon entropy defined as $H(  x )  \equiv-\int p(  x)  \ln p( x)  d x$ for a continuous random variable $x$ with the probabilistic distribution $p (x )$. The results are plotted in Figs.~\ref{fig3}(a) and \ref{fig3}(b). As expected, the Shannon entropies for both $\phi/\pi$ and $\theta/\pi$ approach their uniform limit $\ln 2$ at high temperatures. Deviation from $\ln 2$ defines a crossover temperature scale for the onset of non-uniform probabilistic distribution due to either flux or bond  phase correlation. A tentative phase diagram can then be constructed based on the intensity plots in the insets and the amplitude analysis in Fig.~\ref{fig1}(c). As shown in Fig.~\ref{fig3}(c), the dashed line $T_{\theta}$ marks a crossover to the region with non-uniform $p(\theta)$, while the dash-dotted lines, $T_{\phi(\pi)}$ and $T_{\phi(0)}$, mark the crossover to $\pi$ or 0-flux dominated regions, respectively. Note that these lines are not phase transitions but only serve as a tentative guide to the eye. Approaching zero temperature, as shown in Fig.~\ref{fig3}(d), the curves of $p(\phi)$ at $\phi=\pi$ and $\phi=0$ as a function of $J_{\text{K}}$ cross each other exactly at the transition between regions II and III, indicating that the flux phase $\phi$ distributes uniformly and restores its full symmetry at this point. The phase diagram is therefore divided tentatively into five regions: region I is dominated by $\pi$-flux; region II is a crossover with coexisting $\pi$-flux and hybridization (bond phase) correlation; regions III and IV are dominated mostly by hybridization; region V contains weakly coupled local spins and almost decoupled conduction electrons. To exclude possible finite size effect, we have performed calculations on a $24 \times 24$ lattice using the traveling cluster approximation method \cite{Kumar2006EPJB}, and the results confirmed all five regions.

\begin{figure}[ptb]
\begin{center}
\includegraphics[width=8.6cm]{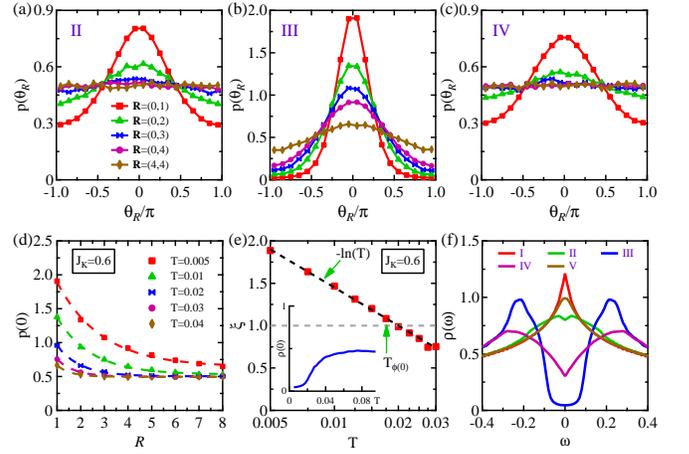}
\end{center}
\caption{The probabilistic distribution $p(\theta_{R})$ for a shortest path between two end sites with $\textbf{R}=( R_x,R_y )$ in regions (a) II, (b) III, and (c) IV. The distribution is always uniform in regions I and V and therefore not shown. (d) The probability $p(\theta_{R}=0)$ as a function of the path length $R$ ($\equiv\vert R_x\vert+\vert R_y\vert$) for different temperatures at $J_{\text{K}}=0.6$. The dashed line is the fit using $A \operatorname{e}^{-R/\xi} +B$. (e) The extracted correlation length $\xi$ as a function of temperature at $J_{\text{K}}=0.6$. We find $\xi$ follows a logarithmic temperature dependence (dashed line) and is roughly one near $T_{\phi(0)}$. The conduction electron density of states at $\omega=0$ (inset) shows that the Kondo insulating (or indirect hybridization) gap opens roughly below the same temperature, with a precursor pseudogap state at higher temperatures. (f) Comparison of typical conduction electron density of states in different regions of the phase diagram. The parameters are chosen as $T=0.005$, $J_{\text{K}}=0.1$ for region I; $T=0.005$, $J_{\text{K}}=0.35$ for II; $T=0.005$, $J_{\text{K}}=0.6$ for III; $T=0.03$, $J_{\text{K}}=0.6$ for IV; $T=0.08$, $J_{\text{K}}=0.1$ for V.}
\label{fig4}
\end{figure}

To clarify the hybridization properties in regions II-IV, we extend the definition of the bond phase to
\begin{align}
\theta_{R}&\equiv\theta_{i_0 i_1}+\theta_{i_1 i_2}+\cdots+\theta_{i_{R-1}i_R}\,\, \text{mod}\,\, 2\pi  \nonumber\\
&=\operatorname*{Im}\ln\left( V_{i_0}\chi_{i_0 i_1}\overline{V}_{i_1} V_{i_1}\chi_{i_1 i_2}\cdots \chi_{i_{R-1}i_R}\overline{V}_{i_R}\right),
\end{align}
where $i_0i_1i_2...i_R$ denotes a path of length $R$ linking two end sites at $\textbf{r}_{i_0}$ and $\textbf{r}_{i_R}\equiv\textbf{r}_{i_0}+\textbf{R}$. Since $\overline{V}_{j} V_{j}=\vert V_{j}\vert^2$ do not contribute a phase, we have also $\theta_{R}=\operatorname*{Im}\log( V_{i_0}\chi_{i_0 i_1}\chi_{i_1 i_2}\cdots \chi_{i_{R-1}i_R}\overline{V}_{i_R})$, which is a gauge-invariant quantity describing phase correlation of the hybridization fields on two end sites mediated by inter-site magnetic correlations along the path. Figures~\ref{fig4}(a)-\ref{fig4}(c) compare the distribution $p ( \theta_R )$ for different $R$ in regions II, III, and IV, respectively. We find it rapidly decays to uniform distribution with increasing $R$ in II and IV but remains peaked at $\theta_R=0$ in region III.

To quantify this decay, we plot in Fig.~\ref{fig4}(d) the probability $p(\theta_R=0)$ at $J_{\text{K}}=0.6$ as a function of the length $R$ for the shortest path linking two sites and fit the curves with an exponential function $p(\theta_R=0)  =A \operatorname{e}^{- R/\xi} +B$ (dashed lines). This allows us to extract a characteristic correlation length $\xi$ which reflects an effective spatial extension of the influence of the hybridization at one site to reach other sites on the Kondo lattice. As shown in Fig.~\ref{fig4}(e), $\xi$ is less than $1$ (in the unit of lattice parameter) at high temperatures but increases logarithmically (dashed line) with lowering temperature. The latter seems to be consistent with the phenomenological two-fluid model \cite{Nakatsuji2004PRL,Curro2004PRB,Yang2008PRL}. Remarkably, the temperature where $\xi \approx 1$ agrees roughly with $T_{\phi(0)}$, the crossover between regions III and IV estimated from the Shannon entropy. Thus, region IV (and II) represents a state where the phase correlation is developed only on short range between nearest-neighbor sites, while in region III, the hybridization fields start to extend their influence in space and build a long-range phase coherence on the lattice.

A direct consequence of the phase correlation may be found on the conduction electron density of states $\rho( \omega)$ calculated using Eq.~(\ref{eq2}). A twisted boundary condition $c_{j}^{\dag}\rightarrow c_{j}^{\dag}\operatorname{e}^{\operatorname*{i} \bm{\psi} \cdot \mathbf{r}_{j}}$ was used to reduce the finite size effect and obtain a smooth curve \cite{Gros1996PRB,Li2018PRL}. The results are presented in Fig.~\ref{fig4}(f) after being averaged over $20 \times 20$ twisted boundary configurations of $\bm{\psi}=\left(  \psi_{x},\psi_{y}\right)$ with both $\psi_{x}$ and $\psi_{y}$ regularly spaced in $\left[  0,\frac{2\pi}{\sqrt{N_{0}}}\right)$. We find that the conduction electron spectra are barely affected in regions I and V, while a pseudogap is developed in regions II and IV where the hybridization bond phase is short-range correlated. This is associated with the ARPES band bending due to the opening of direct hybridization gap \cite{Chen2017PRB}. Only in region III, we see a fully opened (indirect hybridization) gap with a small remaining spectral weight around $\omega \approx 0$ due to Lorentzian broadening ($\delta=0.01$) used in the calculations. The full gap opening temperature is in rough accordance with the growth of $\xi$ in the bond phase distribution, as compared in the inset of Fig.~\ref{fig4}(e) for $J_{\text{K}}=0.6$. Thus, the heavy electron emergence or Kondo insulating state is closely related to the development of long-range phase coherence of hybridization fields mediated by inter-site magnetic correlations, suggesting the importance of our defined bond phase in describing heavy fermion physics beyond the usual mean-field picture and local approximations. The distinction of short- and long-range phase correlations provides a potential microscopic explanation of the two-stage scenario for the pump-probe experiment \cite{Liu2020PRL,Pei2021PRB} and lays a theoretical basis for resolving the conflict between ARPES and transport data \cite{Chen2017PRB}.

To summarize, we propose a novel scheme to study the hybridization physics of Kondo lattice systems based on static auxiliary field approximation. A gauge-invariant hybridization bond phase is introduced to investigate the spatial phase correlation of the hybridization fields. Its probabilistic distribution allows us to define a characteristic length scale which diverges logarithmically with lowering temperature. A precursor pseudogap state with short-range phase correlation is revealed, before the Kondo insulating (or heavy electron) state emerges when a long-range phase coherence starts to develop. This provides a possible microscopic support of the two-stage hybridization scenario suggested by recent experiment. Our work proposes a novel framework based on spatial phase correlation beyond conventional mean-field and local pictures, and offers a useful tool to explore phase-related physics in Kondo lattice systems.

We acknowledge useful discussions with Yin Zhong. This work was supported by the National Natural Science Foundation of China (NSFC Grants No. 12204075, No. 11974397, No. 12174429, and No. 12147102), the National Key Research and Development Program of MOST of China (Grant No. 2017YFA0303103), and the Strategic Priority Research Program of the Chinese Academy of Sciences (Grant No. XDB33010100).

\end{document}


\title{Development of long-range phase coherence on the Kondo lattice\\
\vspace{0.2cm}
- Supplemental Material -}
\author{Jian-Jun Dong}
\affiliation{Department of Physics and Chongqing Key Laboratory for Strongly Coupled Physics, Chongqing University, Chongqing 401331, China}
\affiliation{Beijing National Laboratory for Condensed Matter Physics and Institute of Physics, Chinese Academy of Sciences, Beijing 100190, China}
\affiliation{University of Chinese Academy of Sciences, Beijing 100049, China}
\author{Yi-feng Yang}
\affiliation{Beijing National Laboratory for Condensed Matter Physics and Institute of Physics, Chinese Academy of Sciences, Beijing 100190, China}
\affiliation{University of Chinese Academy of Sciences, Beijing 100049, China}
\affiliation{Songshan Lake Materials Laboratory, Dongguan, Guangdong 523808, China}

\maketitle

\renewcommand\thesection{\arabic{section}}
\renewcommand\thefigure{S\arabic{figure}}
\renewcommand\theequation{S\arabic{equation}}
\setcounter{figure}{0}
\setcounter{equation}{0}

\subsection{I. The effective action}
\label{appA}

To derive the effective action of the Kondo-Heisenberg model, we first use the Abrikosov pseudofermion representation $\mathbf{S}_{i}=\sum_{\eta\gamma}f_{i\eta}^{\dag}\frac{\bm{\sigma}_{\eta\gamma}}{2}f_{i\gamma}$ to rewrite the Kondo and Heisenberg terms as
follows:
\begin{align}
H_{\text{K}}  & =J_{\text{K}}\sum_{i}\mathbf{s}_{i}\cdot\mathbf{S}_{i}\nonumber\\
& =\frac{J_{\text{K}}}{4}\sum_{i}c_{i\alpha}^{\dag}c_{i\alpha}-\frac{J_{\text{K}}}{2}\sum_{i}c_{i\alpha}^{\dag}c_{i\beta}f_{i\alpha}f_{i\beta }^{\dag},\nonumber\\
H_{\text{H}}  & =J_{\text{H}}\sum_{\left\langle ij\right\rangle } \mathbf{S}_{i}\cdot\mathbf{S}_{j}\nonumber\\
& =\frac{N_{0}J_{\text{H}}}{2}-\frac{J_{\text{H}}}{2}\sum_{\left\langle ij\right\rangle }f_{i\alpha}^{\dag}f_{i\beta}f_{j\alpha}f_{j\beta}^{\dag},
\end{align}%
where $N_{0}$ is the number of lattice sites. The first term of $H_{\text{K}}$ can be absorbed into the definition of the electron chemical potential and the constant term of $H_{\text{H}}$ can be dropped for simplicity. Upon Hubbard-Stratonovich transformations, the above interaction terms can be decomposed as
\begin{align}
H_{\text{K}}  & \rightarrow\frac{J_{\text{K}}}{2}\sum_{i}\left\vert V_{i}\right\vert ^{2}+\frac{J_{\text{K}}}{2}\sum_{i\sigma}\left( V_{i}c_{i\sigma}^{\dag}f_{i\sigma}+\text{H.c.}\right)  ,\nonumber\\
H_{\text{H}}  & \rightarrow\frac{J_{\text{H}}}{2}\sum_{\left\langle ij\right\rangle }\left\vert \chi_{ij}\right\vert ^{2}+\frac{J_{\text{H}}} {2}\sum_{\left\langle ij\right\rangle \sigma}\left(  \chi_{ij}f_{i\sigma}^{\dag}f_{j\sigma}+\text{H.c.}\right).
\end{align}
We then obtain a Lagrangian $\mathcal{L}=\mathcal{L}_{\text{c}}\mathcal{+L}_{\text{f}}\mathcal{+L}_{\text{K}}\mathcal{+L}_{\text{H}}$ with
\begin{align}
\mathcal{L}_{\text{c}}  & =\sum_{i\sigma}c_{i\sigma}^{\dag}\left(\partial_{\tau}+\mu\right)  c_{i\sigma}-t\sum_{\left\langle ij\right\rangle\sigma}\left[  c_{i\sigma}^{\dag}c_{j\sigma}+\text{H.c.}\right]  ,\nonumber\\
\mathcal{L}_{\text{f}}  & =\sum_{i\sigma}f_{i\sigma}^{\dag}\left(\partial_{\tau}+\lambda_{i}\right)  f_{i\sigma}-\sum_{i}\lambda_{i},\nonumber\\
\mathcal{L}_{\text{K}}  & =\frac{J_{\text{K}}}{2}\sum_{i}\left\vert V_{i}\right\vert ^{2}+\frac{J_{\text{K}}}{2}\sum_{i\sigma}\left[ V_{i}c_{i\sigma}^{\dag}f_{i\sigma}+\text{H.c.}\right]  ,\nonumber\\
\mathcal{L}_{\text{H}}  & =\frac{J_{\text{H}}}{2}\sum_{\left\langle ij\right\rangle }\left\vert \chi_{ij}\right\vert ^{2}+\frac{J_{\text{H}}} {2}\sum_{\left\langle ij\right\rangle \sigma}\left[  \chi_{ij}f_{i\sigma }^{\dag}f_{j\sigma}+\text{H.c.}\right]  .
\end{align}%
Here, $c_i$ and $f_i$ are Grassmann variables, $V_i$ and $\chi_{ij}$ are complex variables, and $\lambda_{i}$ is the Lagrange multiplier for the constraint $\sum_{\sigma}f_{i\sigma}^{\dag}f_{i\sigma}=1$.

The partition function is
\begin{equation}
Z=\int\mathcal{D}\left[  c_{i},f_{i},\lambda_{i},V_{i},\chi_{ij}\right] \exp\left[  -\int_{0}^{\beta}d\tau\mathcal{L}\left(  \tau\right)  \right]  ,
\end{equation}%
which faithfully represents the original Kondo-Heisenberg model if the functional integrals are performed exactly. We have then the action $S=\int_{0}^{\beta}d\tau\mathcal{L}\left(  \tau\right)$ in the frequency space,
\begin{align}
S  &  =\sum_{i,l}\frac{\beta J_{\text{K}}\left\vert V_{i,l}\right\vert ^{2}}{2}+\sum_{\left\langle ij\right\rangle ,l}\frac{\beta J_{\text{H}}\left\vert \chi_{ij,l}\right\vert ^{2}}{2}-\sum_{i}\beta\lambda_{i}\nonumber\\
&  \quad+\sum_{nm\sigma}\Psi_{n\sigma}^{\dag}\left(  O_{nm}-\operatorname*{i} \omega_{n}\delta_{nm}\right)  \Psi_{m\sigma},
\end{align}
where $\Psi_{n\sigma}=\left[c_{1\sigma n},\cdots,c_{N_{0}\sigma n},f_{1\sigma n},\cdots,f_{N_{0}\sigma n}\right]  ^{T}$, $O_{nm}$ is the corresponding matrix, and the subscripts $l$ ($n/m$) denote bosonic (fermionic) Matsubara frequency.

The static approximation assumes time-independent auxiliary field variables: $V_{i}\left(  \tau\right)  =V_{i}$ and $\chi_{ij}\left(  \tau\right)  =\chi_{ij}$. In frequency space, this implies $V_{i,n-m}=V_{i}\delta_{nm}$ and $\chi_{ij,n-m}=\chi_{ij}\delta_{nm}$. Hence $O_{nm}=O\delta_{nm}$, with the block form $O=\left[ \begin{array} [c]{cc} T^{c} & M \\ M^{\dag} & T^{f} \end{array} \right]$, where $T^{c}$ ($T^{f}$) is the $N_{0}\times N_{0}$ hopping matrix of conduction electrons (pseudofermions), and $M_{ij}=\delta_{ij}J_{\text{K}}V_i/2$ is a diagonal matrix for their onsite hybridization. Using the integral formula,
\begin{equation}
\int\mathcal{D}\left[  \Psi^{\dag},\Psi\right]  \exp\left(  -\Psi^{\dag}M\Psi\right)  =\det M,
\end{equation}
we can integrate out all fermionic degrees of freedom and obtain the following effective action,
\begin{align}
S_{\text{eff}}  &  =\sum_{i}\frac{\beta J_{\text{K}}\left\vert V_{i} \right\vert ^{2}}{2}+\sum_{\left\langle ij\right\rangle }\frac{\beta J_{\text{H}}\left\vert \chi_{ij}\right\vert ^{2}}{2}-\sum_{i}\beta\lambda_{i}\nonumber\\
&  \quad-2\sum_{n}\ln\det\left(  O-\operatorname*{i}\omega_{n}\right).
\end{align}
The partition function is now
\begin{equation}
Z=\int\mathcal{D}\left[  V_{i},\chi_{ij},\lambda_i\right]  \exp\left(  -S_{\text{eff} }\right), \label{appA_partition}
\end{equation}%
which is an $8N_0$-dimensional integral of the auxiliary fields defined on the lattice.

\subsection{II. The uniform mean-field theory}
\label{appB}
The uniform mean-field theory assumes a real solution,
\begin{equation}
  V_{i}=\overline{V}_{i}=V, \quad \chi_{ij}=\overline{\chi}_{ij}=\chi, \quad \lambda_i=\lambda
\end{equation}
which turns the high-dimensional integral into a three-dimensional one, $Z=\int\mathcal{D}\left[  V,\chi, \lambda \right] \operatorname{e}^{-S_{\text{eff}}}$. Their mean-field values can be derived from the saddle-point approximation by minimizing the free energy, $F=S_{\text{eff}} / \beta$. From $\partial F/\partial\chi=0$, $\partial F/\partial V=0$, and $\partial F/\partial \lambda=0$, we obtain the self-consistent equations:
\begin{align}
\chi &  =-\frac{1}{N_{0}}\sum_{\mathbf{k,\eta=\pm}}\left(  \cos k_{x}+\cos k_{y}\right)  \frac{\eta f\left(  E_{\mathbf{k,}\eta}\right)  \left( E_{\mathbf{k,}\eta}-\xi_{c\mathbf{k}}\right)  }{E_{\mathbf{k,+}} -E_{\mathbf{k,-}}},\nonumber\\
V &  =-\frac{J_{\text{K}}V}{N_{0}}\sum_{\mathbf{k}}\left[  \frac{f\left( E_{\mathbf{k,+}}\right)  -f\left(  E_{\mathbf{k,-}}\right)  }{E_{\mathbf{k,+} }-E_{\mathbf{k,-}}}\right]  ,\nonumber\\
1 &  =\frac{2}{N_{0}}\sum_{\mathbf{k,}\eta\mathbf{=\pm}}\frac{\eta f\left( E_{\mathbf{k,\eta}}\right)  \left(  E_{\mathbf{k,}\eta}-\xi_{c\mathbf{k} }\right)  }{E_{\mathbf{k,+}}-E_{\mathbf{k,-}}},
\end{align}
where $f\left(  x\right)  =1/\left(  1+\operatorname{e}^{\beta x}\right)$ is the Fermi distribution function and
\begin{align}
E_{\mathbf{k,\pm}}  & =\frac{\left(  \xi_{c\mathbf{k}}+\xi_{f\mathbf{k} }\right)  \pm\sqrt{\left(  \xi_{c\mathbf{k}}-\xi_{f\mathbf{k}}\right)^{2}+J_{\text{K}}^{2}V^{2}}}{2},\nonumber\\
\xi_{c\mathbf{k}}  & =-2t\left(  \cos k_{x}+\cos k_{y}\right)  ,\nonumber\\
\xi_{f\mathbf{k}}  & =\chi J_{\text{H}}\left(  \cos k_{x}+\cos k_{y}\right) +\lambda.
\end{align}%
The particle-hole symmetry requires $\lambda=0$.

\subsection{III. The Monte Carlo algorithm}
\label{appC}
We use the usual Metropolis algorithm for the Monte Carlo sampling of the probability distribution function,
\begin{equation}
p\left(  V_{i},\chi_{ij}\right)  =\frac{1}{Z}\exp\left(  -S_{\text{eff} }\right)  , \label{appCdistribution}
\end{equation}%
where $Z$ serves as the normalization factor and $\lambda_i$ have been approximated by their mean-field value $\lambda_i=0$ for simplicity. The Monte Carlo procedures involve:

(1) Starting with a random set of configurations
\begin{equation}
\Phi\equiv\left(  V_{1}^{x},V_{1}^{y},\ldots,V_{N_{0}}^{x},V_{N_{0}}^{y} ,\chi_{1}^{x},\chi_{1}^{y},\ldots,\chi_{2N_{0}}^{x},\chi_{2N_{0}}^{y}\right),
\end{equation}%
where we have given a new label (subscript) for the auxiliary variables in the numerical algorithm.

(2) For each Monte Carlo sweep, we update sequentially one of the $6N_0$ variables in $\Phi$, so that $\Phi_{l}^{\prime}=\Phi_{l}+\Delta\times x$ where $x$ is a random number between $-1$ and $1$ and $\Delta>0$ is the step length to adjust the accepted probability. Then we calculate the change of the effective action
\begin{equation}
\Delta S=S_{\text{eff}}\left[  \Phi^{\prime}\right]  -S_{\text{eff}}\left[\Phi\right].
\end{equation}%
Following the Metropolis algorithm, the proposed configuration $\Phi_{l}^{\prime}$ is accepted with the probability $\min\left[  1,\operatorname{e}^{-\Delta S}\right]$. This procedure is repeated sequentially for all $6N_0$ variables in $\Phi$.

(3) Repeating step (2) to thermalize the system.

(4) After thermalization, 200000 Monte Carlo sweep are performed for the Monte Carlo sampling.

(5) Taking one sample (configuration) every 10 sweeps to reduce self-correlations in the data and calculating the physical quantities by average over a total number of 20000 configurations. An illustration of all samples for the local $V_i$ can be found in Fig. 1(d), from which the marginal distribution of $p(V_i)$ can be obtained.

\subsection{IV. The probabilistic distribution of the amplitudes $|F_i|$ and $|B_{ij}|$}
\label{appD}

\begin{figure}[h]
\begin{center}
\includegraphics[width=8.6cm]{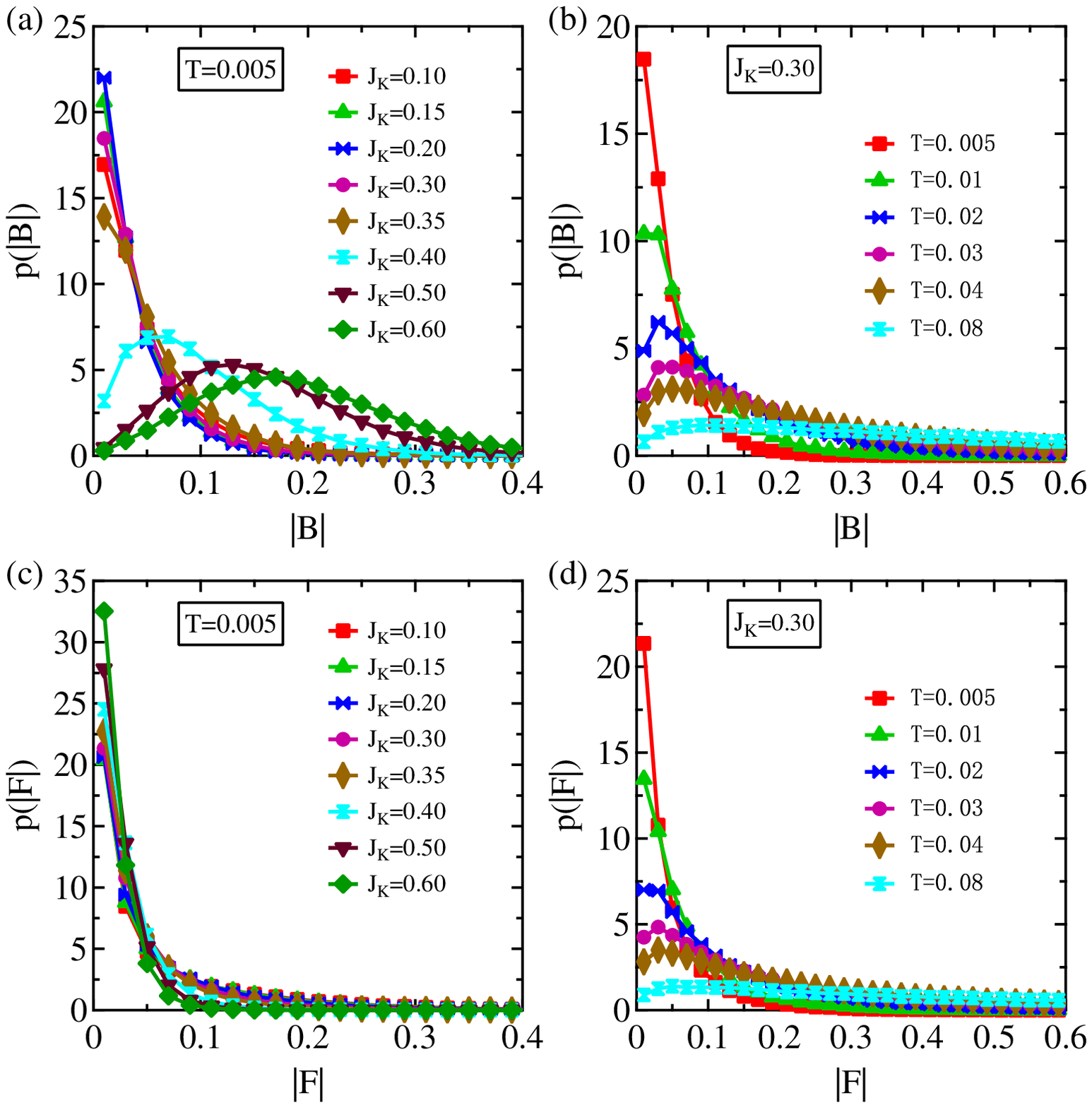}
\end{center}
\caption{Probabilistic distribution of (a)(b) $p(|B|)$ and (c)(d) $p(|F|)$ with varying Kondo coupling $J_{\text{K}}$ or temperature $T$.
}
\label{figSM1}%
\end{figure}

For completeness, we plot in Fig. \ref{figSM1} the probabilistic distribution of the magnitude $p(|B|) \equiv p(|B_{ij}|)$ and $p(|F|) \equiv p(|F_{i})|$ with varying Kondo coupling $J_{\text{K}}$ or temperature $T$. We see a peak in $p(|B|)$ near $|B|=0$ at small $J_{\text{K}}$ or high $T$, which shifts to finite $|B|$ for large $J_{\text{K}}$ at low $T$, reflecting that of $p(|V|)$. The distribution $p(|F|)$ always shows a peak near zero at low $T$ and becomes suppressed at high $T$, reflecting the increase of the magnetic correlation ($|\chi|$ field) with lowering temperature. These results are consistent with the overall picture derived in the main text.

